\documentclass[twocolumn,showpacs,preprintnumbers,amsmath,amssymb,nofootinbib,floatfix]{revtex4}

\usepackage[dvips]{graphicx}
\usepackage{dcolumn}
\usepackage{bm}

\def\mapgeq{\mathbin{\lower.3ex\hbox{$\buildrel>\over{\smash{\scriptstyle\sim}\vphantom{_x}}$}}}
\def\mapleq{\mathbin{\lower.3ex\hbox{$\buildrel<\over{\smash{\scriptstyle\sim}\vphantom{_x}}$}}}
\def\mapgeqeq{\mathbi{\lower.3ex\hbox{$\buildrel>\over{\smash{\scriptstyle\approx}\vphantom{_2}}$}}}
\def\mapleqeq{\mathbin{\lower.3ex\hbox{$\buildrel<\over{\smash{\scriptstyle\approx}\vphantom{_2}}$}}}
\mathchardef\hanaO="724F

\def\Journal#1#2#3#4{{#1} {\bf #2} (#4) #3}

\def\NPB{Nucl. Phys. B}

\def\NPSUPPL{Nucl. Phys. Proc. Suppl.}
\def\PLB{{Phys. Lett.} B}

\def\PLBOLD{Phys. Lett.}
\def\PRL{Phys. Rev. Lett.}
\def\RMP{Rev. Mod. Phys.}
\def\PRD{Phys. Rev. D}

\def\PTP{Prog. Theor. Phys.}
\def\JHEP{JHEP}

\def\EPJ{Euro. Phys. J. C}

\def\JETPUSSR{Sov. Phys. JETP}

\def\ZETP{Zh. Eksp. Teor. Fiz.}

\def\IJMPE{Int. J. Mod. Phys. E}

\def\SCI{Science}
\def\APJ{Astrophys. J.}

\def\NJP{New J. Phys.}

\def\Erratum{Erratum-ibid}

\begin{document}


 \title{Bi-pair Neutrino Mixing}

\author{Teruyuki Kitabayashi}
\email{teruyuki@keyaki.cc.u-tokai.ac.jp}
\author{Masaki Yasu\`{e}}%
\email{yasue@keyaki.cc.u-tokai.ac.jp}
\affiliation{\vspace{3mm}%
\sl Department of Physics, Tokai University,\\
4-1-1 Kitakaname, Hiratsuka, Kanagawa 259-1292, Japan\\
}

\date{January, 2011}

\begin{abstract}
A new type of neutrino mixing named bi-pair neutrino mixing is proposed to describe the current neutrino mixing pattern with a vanishing 
reactor mixing angle and is determined by 
a mixing matrix with two pairs of identical magnitudes of matrix elements.  As a result, we predict 
$\sin^2\theta_{12}=1-1/\sqrt{2}\left(\approx 0.293\right)$ for the solar neutrino mixing
and either $\sin^2\theta_{23}=\tan^2\theta_{12}$ or $\cos^2\theta_{23}=\tan^2\theta_{12}$ for the atmospheric neutrino mixing.
We determine flavor structure of a mass matrix $M$, leading to diagonal masses of $m_{1,2,3}$, 
 and find that
$\left| M_{\mu \mu } - M_{ee}/t_{12}^2 \right|:\left| {{M_{\mu \tau }}} \right|:\left| M_{\tau \tau } - M_{ee}/t_{12}^2  \right| $
=$t^2_{23}:\left| t_{23} \right|:1$ 
for the normal mass hierarchy if $m_1=0$, where $t_{ij}$=$\tan\theta_{ij}$ ($i,j$=$1,2,3$) and $M_{ij}$ ($i,j$=$e,\mu,\tau$) 
stand for flavor neutrino masses.  For the inverted mass hierarchy,
the bi-pair mixing scheme turns out to satisfy the strong scaling ansatz requiring that
$\left| {{M_{\mu \mu }}} \right|:\left| {{M_{\mu \tau }}} \right|:\left| {{M_{\tau \tau }}} \right| = 1:\left|t_{23}\right|:t_{23}^2$ 
if $m_3=0$.
\end{abstract}

\maketitle
More than ten years have passed since the first confirmation of the neutrino oscillations by the Super-Kamiokande collaboration, 
who observed the oscillation of atmospheric neutrinos \cite{SK}. Subsequent experimental observations have also confirmed 
the solar neutrino oscillation \cite{OldSolor,Sun}.  These oscillations really occur in terrestrial neutrinos \cite{K2K,Reactor}.
The other oscillation arising from reactor neutrino mixing has not yet been observed \cite{Theta13}.  
Theoretically, mixings of three flavor neutrinos $\nu_{e,\mu,\tau}$ can account for the neutrino oscillations
if neutrinos are massive and are characterized by three mixing angles $\theta_{12,23,13}$ associated with the mixings of 
$\nu_e$-$\nu_\mu$, $\nu_\mu$-$\nu_\tau$ and $\nu_e$-$\nu_\tau$, respectively.
The Pontecorvo-Maki-Nakagawa-Sakata (PMNS) unitary matrix \cite{PMNS} parameterized by these mixing angles 
is used to convert mass eigenstates of neutrinos into the flavor neutrinos. 

The observed results of the mixing angles are summarized as \cite{NuData}:
\begin{eqnarray}
\sin ^2 \theta _{12} & = &0.304
{\footnotesize
{\begin{array}{*{20}c}
   { + 0.022}  \\
   { - 0.016}  \\
\end{array}}
}~(0.27-0.35),
\nonumber \\
\sin ^2 \theta _{23} & = &0.50
{\footnotesize
{\begin{array}{*{20}c}
   { + 0.07}  \\
   { - 0.06}  \\
\end{array}}
}~(0.39-0.63),
\nonumber \\
\sin ^2 \theta _{13} & = &0.01
{\footnotesize
 {\begin{array}{*{20}c}
   { + 0.016}  \\
   { - 0.011}  \\
\end{array}}
}~(\leq 0.04),
\label{Eq:NuDataAngle}
\end{eqnarray}
for the 1$\sigma$ range, where the values in the parentheses denote the 2$\sigma$ range.
There is a theoretical prediction of these mixing angles based on the tri-bimaximal mixing scheme \cite{tribimaximal}, 
which yields
\begin{eqnarray}
\sin^2\theta_{12}=\frac{1}{3},
\quad
\sin^2\theta_{23}=\frac{1}{2},
\quad
\sin^2\theta_{13}=0.
\label{Eq:Tribimaximal}
\end{eqnarray}
These predictions are consistent with the 2$\sigma$ data although $\sin^2\theta_{12}$ slightly exceeds 
the allowed range of the 1$\sigma$ data. 

In this short note, we would like to find new mixing schemes \cite{Others}, 
which may well describe the solar neutrino mixing.  
To do so, we demand that at least 
two of the magnitudes of matrix elements be equal to each other as in the tri-bimaximal case, which shows
\begin{eqnarray}
&&
\left|U_{12}\right|=\left|U_{22}\right|=\left|U_{32}\right|,
\nonumber \\
&&
\left|U_{21}\right|=\left|U_{31}\right|,
\nonumber \\
&&
\left|U_{23}\right|=\left|U_{33}\right|.
\label{Eq:Tribimaximal-elements}
\end{eqnarray}
We have parameterized the PMNS mixing matrix $U$ to be $U=K^0U^0$:
\begin{eqnarray}
U^0&=&\left( \begin{array}{ccc}
  c_{12} &  s_{12}&  0\\
  -c_{23}s_{12}&  c_{23}c_{12}&  s_{23}\\
  s_{23}s_{12} &  -s_{23}c_{12}& c_{23}\\
\end{array} \right),
\nonumber \\
K^0 &=& {\rm diag}(1, e^{i\phi_2/2}, e^{i\phi_3/2}),
\label{Eq:Uu}
\end{eqnarray}
where $c_{ij}=\cos\theta_{ij}$, $s_{ij}=\sin\theta_{ij}$ ($i,j$=1,2,3) and $\phi_{2,3}$ are CP-violating Majorana phases.  

As long as $\theta_{13}=0$ is maintained, 
it is not difficult to search for alternative relations to Eq.(\ref{Eq:Tribimaximal-elements}) 
for the given values of Eq.(\ref{Eq:NuDataAngle}).  There are only two possibilities, which shows
\begin{eqnarray}
&&
\left|U_{12}\right|=\left|U_{32}\right|,
\nonumber \\
&&
\left|U_{22}\right|=\left|U_{23}\right|,
\label{Eq:bipair-elements-1}
\end{eqnarray}
as the case (1), and 
\begin{eqnarray}
&&
\left|U_{12}\right|=\left|U_{22}\right|,
\nonumber \\
&&
\left|U_{32}\right|=\left|U_{33}\right|,
\label{Eq:bipair-elements-2}
\end{eqnarray}
as the case (2). These equations in turn provide useful relationship among the mixing angles:
\begin{eqnarray}
\frac{\sin\theta_{12}}{\cos\theta_{12}}=\frac{\cos\theta_{12}}{\sqrt{1+\cos^2\theta_{12}}},
\label{Eq:prediction-12}
\end{eqnarray}
as well as
\begin{eqnarray}
\sin^2\theta_{23}=\tan^2\theta_{12},
\label{Eq:prediction-23-1}
\end{eqnarray}
for the case (1), and
\begin{eqnarray}
\cos^2\theta_{23}=\tan^2\theta_{12},
\label{Eq:prediction-23-2}
\end{eqnarray}
for the case (2).  Numerically, these relations predict
\begin{eqnarray}
\sin^2\theta_{12}&=&1-\frac{1}{\sqrt{2}}\approx 0.293,
\nonumber \\
{\sin ^2}{\theta _{23}} &=& \left\{ \begin{array}{l}
 \sqrt 2  - 1 \approx 0.414 \cdots \rm{the~case~(1)}\\ 
 2 - \sqrt 2  \approx 0.586 \cdots \rm{the~case~(2)}\\ 
 \end{array} \right.,
\label{Eq:prediction-numercal}
\end{eqnarray}
which are consistent with the 2$\sigma$ data.  

Our prediction on $\sin^2\theta_{23}$ is slightly inconsistent with
the 1$\sigma$ data as in the similar situation to that of $\sin^2\theta_{12}$ in the tri-bimaximal mixing.  
However, it is well known that the corresponding 2-3 mixing 
in charged leptons (labeled by $\theta_{23}^\ell$) can produce additional contribution 
to $\theta_{23}$ without affecting $\theta_{12}$ and $\theta_{13}$.  Namely, we obtain that
\begin{eqnarray}
\theta_{23}=\theta^\nu_{23}-\theta^\ell_{23},
\label{Eq:12-corrected}
\end{eqnarray}
where $\theta^\nu_{23}$ is given by $\theta_{23}$ in Eq.(\ref{Eq:prediction-numercal}). 
Therefore, if charged leptons have a mass matrix $M_\ell$ described by
\begin{eqnarray}
M_\ell =
\left( {\begin{array}{*{20}{c}}
   {{m_e}} & 0 & 0  \\
   0 & \ast & \ast  \\
   0 & \ast & \ast  \\
\end{array}} \right),
\label{Eq:charged-lepton-mass}
\end{eqnarray}
appropriate correction automatically comes in $\theta_{23}$ so that $\theta_{23}$ can be shifted to the 1$\sigma$ region.  Other 
corrections may arise from the renormalization effect \cite{renormalization} if the bi-pair mixing is generated at a higher scale such as 
the seesaw scale, where the seesaw mechanism \cite{seesaw} gets active.

Having understood that the bi-pair mixing is another candidate predicting the reasonable values of $\theta_{12}$ and $\theta_{23}$,
we discuss its implication of flavor structure of the neutrino masses.  
It has been discussed that any models with $\sin\theta_{13}=0$ should be described by 
the following flavor neutrino mass matrix $M_\nu$ with general phase structure \cite{theta_13=0}
\begin{eqnarray}
&&
M_\nu =
\left(
  \begin{array}{ccc}
    a & e^{i\alpha }\left| b \right| & -t_{23}e^{i\beta }\left| b \right| \\
           & d & e \\
           &        & f \\
  \end{array}
\right),
\nonumber \\
&&
f = e^{4i\gamma}d+e^{2i\gamma}\frac{1-t_{23}^2}{t_{23}}e,
\label{Eq:Mnu-13=0}
\end{eqnarray}
where $a$, $b$, $d$ and $e$ are complex numbers, $\alpha$ and $\beta$ are phases, $\gamma$ is given by
\begin{eqnarray}
&&
\gamma = \frac{\beta-\alpha}{2},
\label{Eq:gamma}
\end{eqnarray}
and $t_{23}=\tan\theta_{23}$.  This result is a consequence of 
the direct calculation of $U^TM_\nu U=M_{\rm diag}$, 
where $M_{\rm diag} = {\rm diag}(m_1,m_2,m_3)$ is the diagonal neutrino mass matrix.  One 
important point is that $U$ should contain redundant phases, which are denoted by $\gamma$ and also by $\rho$ (to be used in Eq.(\ref{Eq:tan2-12})), 
to take care of general phase structure of $M_\nu$ \cite{generalU1,generalU2}.  For $U^{PDG}$ as the 
standard parameterization of $U$ adopted by the Particle Data Group (PDG) \cite{PDG},  $M_\nu$ is shifted to a modified 
mass matrix $M$ after $\rho$ and $\gamma$ present in $U$ are transferred 
to $M$.

Owing to the rephasing ambiguity 
in the charged lepton sector, one can choose three flavor masses to be real numbers, where
$d$ can be taken to be positive without loss of generality. As a result, we obtain
\begin{eqnarray}
&&
M_\nu =
\left(
  \begin{array}{ccc}
    \kappa_a\left| a \right| & e^{i\alpha }\left| b \right| & -t_{23}e^{i\beta }\left| b \right| \\
           & \left| d \right|& \kappa_e\left| e \right|\\
           &        & f \\
  \end{array}
\right),
\nonumber \\
&&
f = e^{4i\gamma}\left| d \right|+\frac{1-t_{23}^2}{t_{23}}\kappa_ee^{2i\gamma}\left| e \right|,
\label{Eq:Mnu-13=0-final}
\end{eqnarray}
where $\kappa_{a,e}$ take care of the sign of $a$ and $e$.
The mixing angles $\theta_{12}$ is given by
\begin{eqnarray}
&&
 \tan 2{\theta _{12}} = \frac{2e^{i\xi}}{c_{23}}
 \frac{\left| b \right|}{e^{2i\gamma}\left| d \right| - {t_{23}}{\kappa _e}\left| e \right| - {\kappa _a}e^{2i\rho }\left| a \right|},
\label{Eq:tan2-12}\\
&&
\xi=\rho+\gamma+\alpha\left(=\rho+\frac{\alpha+\beta}{2}\right),
\label{Eq:xi-tan2-12}
\end{eqnarray}
which determines the phase $\rho$ expressed in terms of flavor neutrino masses for a given value of $\theta_{12}$.  

After redundant phases are removed from $U$, $U$ becomes $U^{PDG}$ and, accordingly, $M_\nu$ is shifted to
\begin{eqnarray}
&&
M
=
\left( {\begin{array}{*{20}{c}}
   e^{2i\rho }\kappa_a\left| a \right| & e^{i\xi}\left| b \right| &  - t_{23}e^{i\xi}\left| b \right|  \\
   {} & {{e^{ 2i\gamma}}\left| d \right|} & {{\kappa _e}\left| e \right|}  \\
   {} & {} & {{e^{-2i\gamma}}f}\\
\end{array}} \right).
\label{Eq:M-PDG}
\end{eqnarray}
We finally reach $M$ given by
\begin{eqnarray}
M&=&
{e^{-i\left( {\alpha-\beta } \right)}}\left| d \right| I
+
\kappa_e\left| e \right|
\left( {\begin{array}{*{20}{c}}
   { - {t_{23}}} & 0 & 0  \\
   0 & 0 & 1  \\
   0 & 1 & {\frac{{1 - t_{23}^2}}{{{t_{23}}}}}  \\
\end{array}} \right)
\nonumber \\
&&
+
{e^{i\left( {\rho  + \frac{{\alpha  + \beta }}{2}} \right)}}\left| b \right|
\left( {\begin{array}{*{20}{c}}
   A & 1 &  - {t_{23}}  \\
   1 & 0 & 0  \\
   { - {t_{23}}} & 0 & 0  \\
\end{array}} \right),
\label{Eq:M-final}\\
A&=&\frac{t_{12}^2 - 1}{c_{23}t_{12}},
\label{Eq:M-final-A}
\end{eqnarray}
where $t_{12}=\tan\theta_{12}$ and 
$I$ is the unit matrix and $\vert a \vert$ in Eq.(\ref{Eq:M-PDG}) is eliminated by Eq.(\ref{Eq:tan2-12})
to yield Eq.(\ref{Eq:M-final}).  
This mass matrix certainly gives $M$ for the tri-bimaximal neutrino mixing if $t^2_{23}=1$ and $t^2_{12}=1/2$
giving $A=-1$.
The bi-pair neutrino mixing gives
$A = -1/\left|t_{23}\right|$ with $t^2_{23}=1/\sqrt{2}$ and $\sin\theta_{23}=\sigma\tan\theta_{12}$ 
for the case (1), where $\sigma$(=$\pm1$) takes care of the sign of $\sin\theta_{23}$, and 
$A = - t^2_{23}$ with $t^2_{23}=\sqrt 2$ and $\cos\theta_{23}=\tan\theta_{12}$ for the case (2).
More transparent flavor structure for the bi-pair neutrino 
mixing can be found if neutrino mass hierarchies are taken into account.  

Neutrino masses are calculated to be
\begin{eqnarray}
{m_1}{e^{ - i{\varphi _1}}} &=& \kappa _ae^{2i\rho }\left| a \right| - \tan {\theta _{12}}\frac{e^{i\xi}\left| b \right|}{c_{23}},
\nonumber\\
{m_2}{e^{ - i{\varphi _2}}} &=& \kappa _ae^{2i\rho }\left| a \right| + \frac{1}{\tan \theta _{12}}\frac{e^{i\xi}\left| b \right|}{c_{23}},
\label{Eq:nu-masses} \\
 {m_3}{e^{ - i{\varphi _3}}} &=& {e^{ - i\left( {\alpha  - \beta } \right)}}\left| d \right| + \frac{1}{{{t_{23}}}}{\kappa _e}\left| e \right|,
\nonumber
\end{eqnarray}
where the CP-violating Majorana phases $\phi_{2,3}$ are given by $\phi_2=\varphi_2-\varphi_1$ and $\phi_3=\varphi_3-\varphi_1$.
Let us consider that neutrinos exhibit either $m_1$=0 leading to the normal mass hierarchy or $m_3$=0 leading to the inverted mass hierarchy 
as in the minimal seesaw model \cite{minimalSeesaw}, where det($M$)=0 is satisfied.  We, then, find that,
for the normal mass hierarchy,
\begin{eqnarray}
M
&=&
\left( {\begin{array}{*{20}{c}}
   {B{e^{i\left( {\alpha  + \beta } \right)}}\left| b \right|} & {{e^{i\left( {\alpha  + \beta } \right)}}\left| b \right|} & { - {t_{23}}{e^{i\left( {\alpha  + \beta } \right)}}\left| b \right|}  \\
   {} & \begin{array}{l}
\frac{Be^{i\left( {\alpha  + \beta } \right)}\left| b \right|}{t_{12}^2} \\ 
 \quad + {t_{23}}{\kappa _e}\left| e \right| \\ 
 \end{array} & {{\kappa _e}\left| e \right|}  \\
   {} & {} & \begin{array}{l}
{\frac{Be^{i\left( {\alpha  + \beta } \right)}\left| b \right|}{{t_{12}^2}}} \\ 
 \quad + \frac{1}{{{t_{23}}}}{\kappa _e}\left| e \right| \\ 
 \end{array}  \\
\end{array}} \right),
\nonumber \\
B&=&\frac{{\tan {\theta _{12}}}}{{{c_{23}}}}=
\left\{ \begin{array}{l}
 \left| t_{23} \right| \cdots \left(\sin\theta_{23}=\sigma\tan\theta_{12} \right) \\ 
 1 \cdots \left(\cos\theta_{23}=\tan\theta_{12}\right) \\ 
\end{array}
\right.,
\label{Eq:nu-flavor-bipair-normal}
\end{eqnarray} 
where $\rho=(\alpha+\beta)/2$ (mod $\pi$) from $m_1=0$ and,
for the inverted mass hierarchy, 
\begin{eqnarray}
M
=
\left( {\begin{array}{*{20}{c}}
   \kappa _a{{e^{2i\rho }}\left| a \right|} & e^{i\xi}\left| b \right| &  - t_{23}e^{i\xi}\left| b \right|  \\
   {} & { - \frac{1}{t_{23}}\kappa _e\left| e \right|} & {{\kappa _e}\left| e \right|}  \\
   {} & {} & { - {t_{23}}{\kappa _e}\left| e \right|}  \\
\end{array}} \right),
\label{Eq:nu-flavor-bipair-inverted}
\end{eqnarray}
where ${e^{-i\left(\alpha-\beta\right)}}\left| d \right| =  - \kappa _e/t_{23}\left| e \right|$ 
from $m_3=0$, thus, leading to $\alpha$=$\beta$ (mod $\pi$) and $\rho$ is determined so as to satisfy 
Eq.(\ref{Eq:tan2-12}), which is used to express this mass matrix in terms of $\left| b \right|$ and $\left| e \right|$. 

We observe that flavor structure of $M$ for the bi-pair neutrino mixing shows
\begin{eqnarray}
&&
 \left| {{M_{e\mu }}} \right|:\left| {{M_{e\tau }}} \right| = 1:\left|t_{23}\right|,
\label{Eq:nu-flavor-bipair-inverted-normal-emu-etau}\\
&&\arg \left( {{M_{e\mu }}} \right) = \arg \left( {{M_{e\tau }}} \right)~ ({\rm mod}~\pi),
\label{Eq:nu-flavor-bipair-inverted-normal-emu-etau-arg}
\end{eqnarray}
and
\begin{itemize}
\item for the normal mass hierarchy,
\begin{eqnarray}
&&
\left| {{M_{ee}}} \right|
=
\left\{ \begin{array}{l}
 \left| {{M_{e\tau }}} \right| \cdots \left(\sin\theta_{23}=\sigma\tan\theta_{12} \right) \\ 
 \left| {{M_{e\mu }}} \right| \cdots \left(\cos\theta_{23}=\tan\theta_{12}\right) \\ 
 \end{array} \right.,
\label{Eq:nu-flavor-bipair-normal-final}\\
&&
\left| M_{\mu \mu } - \frac{M_{ee}}{t_{12}^2} \right|:\left| {{M_{\mu \tau }}} \right|:\left| M_{\tau \tau } - \frac{M_{ee}}{t_{12}^2}  \right| 
\nonumber\\
&&\qquad= t^2_{23}:\left| t_{23} \right|:1, 
\label{Eq:nu-flavor-bipair-normal-final2}\\
&&
\arg \left( {{M_{ee}}} \right) = \arg \left( {{M_{e\mu }}} \right) ~ ({\rm mod}~\pi),
\label{Eq:nu-flavor-bipair-normal-arg}
\end{eqnarray}
\item for the inverted mass hierarchy,
\begin{eqnarray}
&&
 \left| {{M_{\mu \mu }}} \right|:\left| {{M_{\mu \tau }}} \right|:\left| {{M_{\tau \tau }}} \right| = 1:\left|t_{23}\right|:t_{23}^2.
\label{Eq:nu-flavor-bipair-inverted-final}
\end{eqnarray}
\end{itemize}
It is noted that Eq.(\ref{Eq:nu-flavor-bipair-inverted}) may satisfy the strong scaling ansatz \cite{SSA} 
since $\sin\theta_{13}=0$ and $m_3=0$.  It is evident that the resulting mass matrix does 
satisfy the strong scaling ansatz requiring the relation of Eq.(\ref{Eq:nu-flavor-bipair-inverted-final}) 
(as well as Eq.(\ref{Eq:nu-flavor-bipair-inverted-normal-emu-etau})).  
Therefore, when $m_3\neq 0$, the bi-pair neutrino mixing 
provides an example of the approximate strong scaling ansatz, where Eq.(\ref{Eq:nu-flavor-bipair-inverted-final}) is 
approximately satisfied. If $m_1=0$ for the normal mass hierarchy, the relation Eq.(\ref{Eq:nu-flavor-bipair-normal-final2}) 
including $M_{ee}$ can be predicted.
\footnote{If $m^2_3\gg m^2_2$ is further imposed, the condition of 
 $b\approx 0$ leading to $M_{ee}\approx 0$ should be satisfied and Eq.(\ref{Eq:nu-flavor-bipair-normal-final2}) 
 becomes
$\left| M_{\mu \mu }\right|:\left| M_{\mu \tau } \right|:\left| M_{\tau \tau }\right| \approx t^2_{23}:\left| t_{23} \right|:1$.}
 
In summary, we have found that the bi-pair mixing well reproduces the current neutrino mixings and is 
 described by $U_{BP}$:
\begin{eqnarray}
U_{BP}&=&\left( \begin{array}{ccc}
  c_{12} &  s_{12}&  0\\
  -t^2_{12}&  t_{12}&  t_{12}\\
  s_{12}t_{12} &  -s_{12}& t_{12}/c_{12}\\
\end{array} \right),
\label{Eq:bi-pair-1}
\end{eqnarray}
for the case (1), and
\begin{eqnarray}
U_{BP}&=&\left( \begin{array}{ccc}
  c_{12} &  s_{12}&  0\\
  -s_{12}t_{12}&  s_{12}&  t_{12}/c_{12}\\
  t^2_{12} &  -t_{12}& t_{12}\\
\end{array} \right),
\label{Eq:bi-pair-2}
\end{eqnarray}
for the case (2), where $s^2_{12}$ is predicted to be: $s^2_{12}=1-1/\sqrt{2}$. 
The bi-pair mixing scheme turns out to be complementary to the tri-bimaximal mixing scheme
in a sense that 
\vspace{-2mm}
\begin{itemize}
\item the bi-pair mixing predicts $\sin^2\theta_{12}=0.293$, which well describes the solar neutrino mixing 
while it predicts $\sin^2\theta_{23}=0.414/0.586$, which gives slight deviation of the atmospheric neutrino 
mixing angle from the 1$\sigma$ region, and 
\item the tri-bimaximal mixing predicts $\sin^2\theta_{23}=0.5$, which well describes the atmospheric neutrino mixing 
while it predicts $\sin^2\theta_{12}=0.333$, which gives slight deviation of the solar neutrino mixing angle from the 1$\sigma$ region.
\end{itemize}

We have clarified the flavor structure of the neutrino mass matrix giving $\sin\theta_{13}=0$, which is described by Eq.(\ref{Eq:M-final}) 
 as long as the parameterization of $U^{PDG}$ is adopted.  
For the bi-pair mixing, in the simplest cases with $m_1=0$ for the normal mass hierarchy 
and $m_3=0$ for the inverted mass hierarchy,
the phase structure is subject to Eqs.(\ref{Eq:nu-flavor-bipair-inverted-normal-emu-etau-arg}) 
and (\ref{Eq:nu-flavor-bipair-normal-arg}).  
We have also predicted Eq.(\ref{Eq:nu-flavor-bipair-normal-final2}) for the normal mass hierarchy, 
which should be compared with Eq.(\ref{Eq:nu-flavor-bipair-inverted-final}) 
for the strong scaling ansatz valid in the inverted mass hierarchy.

Finally, we point out that the results of Eqs.(\ref{Eq:nu-flavor-bipair-inverted-normal-emu-etau}) 
and (\ref{Eq:nu-flavor-bipair-inverted-normal-emu-etau-arg}) 
for both normal and inverted mass hierarchies
and of Eq.(\ref{Eq:nu-flavor-bipair-normal-final2}) for the normal mass hierarchy 
and Eq.(\ref{Eq:nu-flavor-bipair-inverted-final}) for the inverted mass hierarchy 
are not only valid in the bi-pair mixing scheme and but also 
valid for any models with $\sin\theta_{13}=0$, where $\theta_{12,23}$ are simply regarded as free parameters.  
The bi-pair mixing scheme provides 
a good example of these properties of the flavor neutrino masses.  
We will discuss the detailed feature of our flavor neutrino mass matrix as well as 
phenomenological implication of the bi-pair mixing scheme based on 
Majorana CP violation \cite{Majorana} from Eq.(\ref{Eq:nu-masses}) and on leptogenesis \cite{leptogenesis} 
in a future publication \cite{future}.

\end{document}